\documentclass[runningheads]{llncs}
\usepackage[T1]{fontenc}
\usepackage{graphicx}
\usepackage{times}
\usepackage{epsfig}
\usepackage{graphicx}
\usepackage{amsmath}
\usepackage{amssymb}
\usepackage{hyperref}
\graphicspath{ {./images/} }
\begin{document}
\title{Unrolled Creative Adversarial Network For Generating Novel Musical Pieces}
%
%
\author{Pratik Nag\orcidID{0000-0001-5065-5273}}
\authorrunning{P.Nag.}
%
\institute{ 
\email{pratik.nag@kaust.edu.sa}\\
\url{https://pratiknag.com/}}
\maketitle              
\begin{abstract}
Music generation has emerged as a significant topic in artificial intelligence and machine learning. While recurrent neural networks (RNNs) have been widely employed for sequence generation, generative adversarial networks (GANs) remain relatively underexplored in this domain. 
This paper presents two systems based on adversarial networks for music generation. The first system learns a set of music pieces without differentiating between styles, while the second system focuses on learning and deviating from specific composers' styles to create innovative music. By extending the Creative Adversarial Networks (CAN) framework to the music domain, this work introduces unrolled CAN to address mode collapse, evaluating both GAN and CAN in terms of creativity and variation.

\end{abstract}

\section{Introduction}
Machines have long been envisioned as tools capable of replicating learned behaviors. However, instilling creativity in machines remains a formidable challenge in the fields of machine learning and artificial intelligence. Creativity, a quintessential human trait, is pivotal in tasks such as problem-solving and decision-making. For instance, in self-driving car development, enabling machines to make creative decisions is essential for handling unforeseen road events and preventing accidents.

\begin{figure}[htbp]
    \centering
    \includegraphics[width=\textwidth]{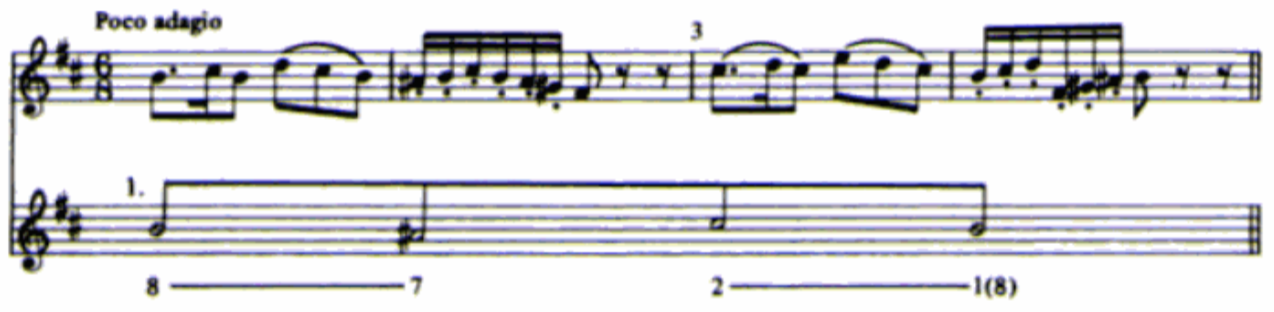}
    \caption{Symphony No.46 in B major, Haydn}
    \label{fig1}
\end{figure}

\begin{figure}[htbp]
    \centering
    \includegraphics[width=\textwidth]{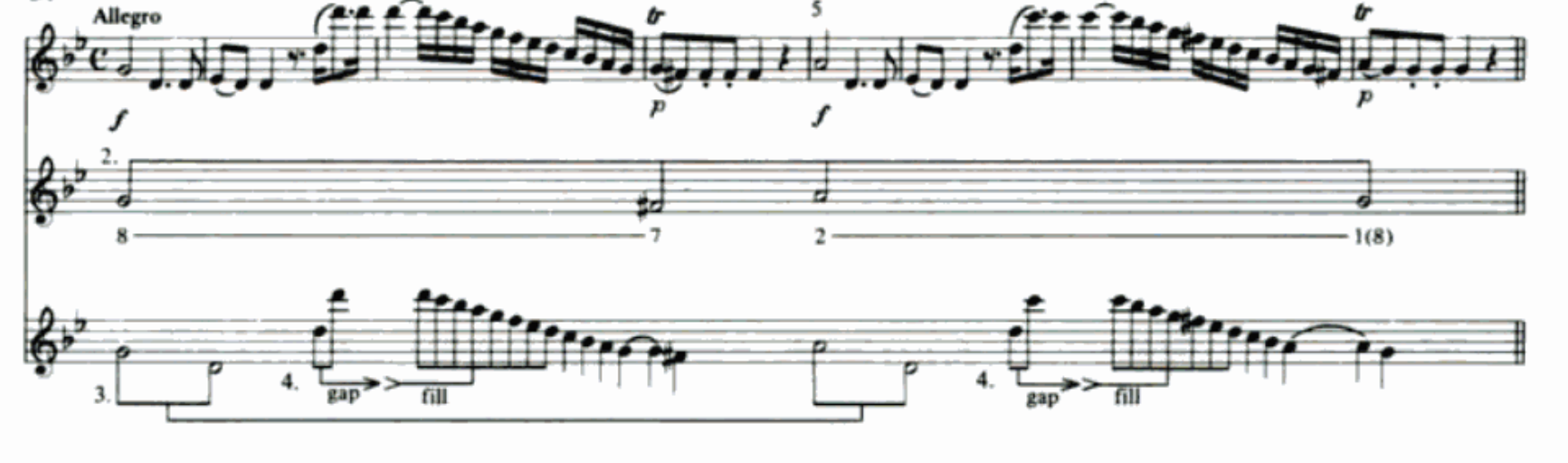}
    \caption{Mozart Piano Quartet in G minor}
    \label{fig2}
\end{figure}

Music generation is particularly complex, involving challenges like learning polyph- ony, rhythm, and intricate patterns of musical notes \cite{muhamed2021symbolic}. Style in music, akin to style in language, is governed by constraints. In language, style emerges from vocabulary selection, grammar, and linguistic preferences. Similarly, in music, patterns of style manifest through pitch choices, rhythms, and dynamics, producing variety within certain constraints \cite{meyer1996style}. For instance, Figures \ref{fig1} and \ref{fig2} illustrate patterns in classical music that replicate specific stylistic elements preferred by renowned composers.

\begin{figure}[!h]
    \centering
    \includegraphics[width=\textwidth]{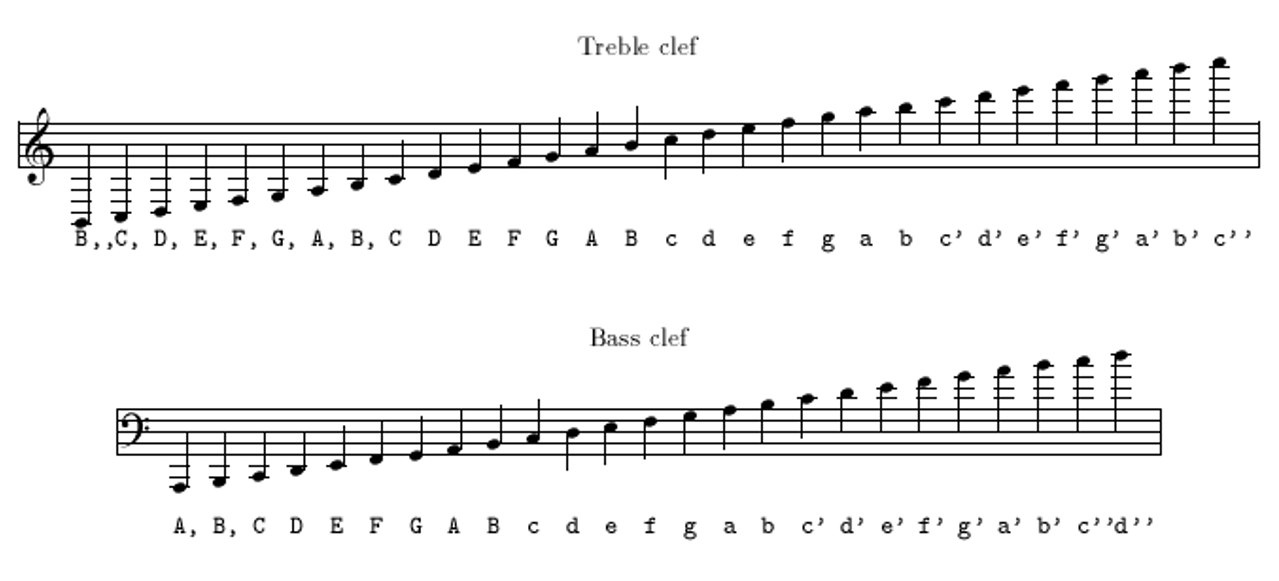}
    \caption{Treble clef and Bass clef}
    \label{fig3}
\end{figure}

\begin{figure}[!h]
    \centering
    \includegraphics[width=\textwidth]{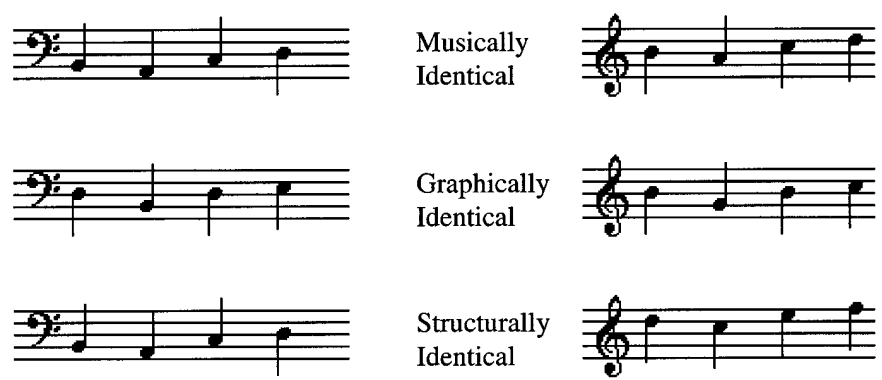}
    \caption{Same-Different exercise example \cite{schon2001naming}}
    \label{fig4}
\end{figure}

Sheet music also presents challenges in interpretation. Notes represented graphically may appear different but sound the same, or conversely, identical-looking sheets may represent different notes due to clef changes \cite{schon2001naming}. Professional musicians rely on experience to identify such nuances naturally. Figures \ref{fig3} and \ref{fig4} demonstrate these complexities, including examples of treble and bass clefs and exercises that highlight the "same-different" note identification challenge.

Recent advancements have explored RNN-based approaches for music generation, focusing on sequence prediction tasks \cite{yang2017midinet,briot2020deep}. On the other hand, research into GAN-based frameworks has been comparatively limited \cite{huang2020pop,davis2014generating,chen2001creating}. Building on these developments, this work adapts the Creative Adversarial Networks (CAN) framework \cite{DBLP} from visual arts to the music domain. By introducing unrolled CAN, this study addresses challenges like mode collapse and evaluates the potential of GAN and CAN in creative music generation.


\section{Technical Details}

\subsection{Generative Adversarial Networks}

Generative Adversarial Networks (GANs), first introduced by Goodfellow \textit{et al.} (2014) \cite{goodfellow2014generative}, consist of two neural networks engaged in a two-player minimax game: a Generator \(G\) and a Discriminator \(D\). The generator maps a noise vector
\[
    z \sim p_{z}(z),
\]
typically drawn from a simple prior distribution (e.g., Gaussian or uniform), to the data space to produce synthetic samples \(G(z)\). The discriminator receives either real data samples
\[
    x \sim p_{\text{data}}(x)
\]
or generated samples \(G(z)\), and outputs a scalar value \(D(x) \in [0,1]\) representing the probability that the input is real.

The minimax value function for the original GAN is defined as
\begin{equation}
    \min_G \max_D V(D,G)
    = \mathbb{E}_{x \sim p_{\text{data}}(x)}[\log D(x)]
    + \mathbb{E}_{z \sim p_z(z)}[\log(1 - D(G(z)))].
\end{equation}

The discriminator attempts to maximize this objective by improving its ability to distinguish real from generated samples, while the generator attempts to minimize it by producing samples that the discriminator classifies as real.

\subsection{Creative Adversarial Networks}

Creative Adversarial Networks (CANs) extend the GAN formulation by incorporating stylistic classification and explicit class ambiguity objectives \cite{DBLP}. In this setup, the discriminator outputs two components:
\begin{itemize}
    \item \(D_r(x)\): the probability that \(x\) is real,
    \item \(D_c(c \mid x)\): the posterior distribution over stylistic classes \(c \in \{1,\dots,K\}\).
\end{itemize}

For real samples, the discriminator is trained both to classify them as real and to correctly identify their style class \(\hat{c}\). For generated samples \(G(z)\), the goal is to enforce \emph{class ambiguity} by maximizing the entropy of the predicted class posterior. This prevents the generator from reproducing known styles and encourages creativity.
The value function for CAN is therefore:
\begin{equation}
\begin{split}
    \min_G \max_D V(D,G)
    =\; &\mathbb{E}_{x,\,\hat{c}\sim p_{\text{data}}}
        \!\left[ \log D_r(x) + \log D_c(\hat{c} \mid x) \right] \\
    &+ \mathbb{E}_{z\sim p_z(z)}
       \Bigg[ \log\!\left(1 - D_r(G(z))\right) \\
    &\qquad - \sum_{k=1}^{K}\!
            \left(
                \frac{1}{K} \log D_c(c_k \mid G(z))
                + \left(1-\frac{1}{K}\right)\!
                   \log(1 - D_c(c_k \mid G(z)))
            \right)
       \Bigg].
\end{split}
\end{equation}
The last term enforces high-entropy outputs over the class distribution, ensuring that the generated samples do not collapse into any known stylistic class.

\subsection{Unrolling Creative Adversarial Networks}

Unrolling Creative Adversarial Networks (Unrolled-CAN) combine the CAN architecture with the unrolled optimization procedure introduced in Unrolled GANs \cite{metz2017unrolled}. The core idea is to allow the generator to anticipate how the discriminator will change in future training steps, thereby smoothing the adversarial dynamics and reducing instances of mode collapse.
In a standard GAN or CAN, parameters are updated as:
\begin{equation}
    G \leftarrow G - \eta \frac{\partial V(D,G)}{\partial G},
\end{equation}
\begin{equation}
    D \leftarrow D - \eta \frac{\partial V(D,G)}{\partial D},
\end{equation}
where \(\eta\) denotes the learning rate.

In an unrolled model, the discriminator is virtually updated for \(k\) unrolling steps (without committing these updates to the actual discriminator). The generator then optimizes the loss with respect to this \(k\)-step unrolled discriminator, denoted \(D^{(k)}\). Thus, the updates become:
\begin{equation}
    G \leftarrow G - \eta \frac{\partial V(D^{(k)},G)}{\partial G},
\end{equation}
\begin{equation}
    D \leftarrow D - \eta \frac{\partial V(D,G)}{\partial D}.
\end{equation}

Here, \(k\) is the number of unrolling steps. Increasing \(k\) allows the generator to account for the discriminator’s future reaction, effectively weakening overly strong discriminators and enabling the generator to explore more diverse solutions.

\begin{figure}[!h]
    \centering
    \includegraphics[width=\textwidth]{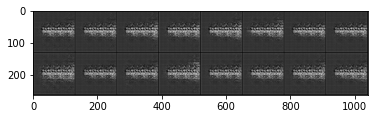}
    \caption{Example of standard CAN output with $k=0$, showing convergence to nearly identical samples after 20 epochs.}
    \label{fig12}
\end{figure}

\begin{figure}[!h]
    \centering
    \includegraphics[width=\textwidth]{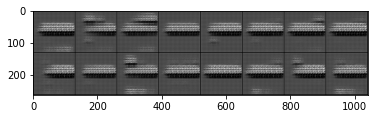}
    \caption{Example of Unrolled-CAN output with $k=1$, showing increased variability and reduced mode collapse after 20 epochs.}
    \label{fig13}
\end{figure}

Figures~\ref{fig12} and \ref{fig13} illustrate the difference between a standard CAN (\(k=0\)) and an Unrolled-CAN (\(k=1\)), showing improved diversity and reduced convergence to a single mode after 20 epochs.

\section{Methodology}

In this section the methodology of music GAN, CAN and unrolled CAN pipeline is discussed. Starting with the data sets used, data preprocessing technique, and the model architecture. 
\subsection{Dataset}
Three data sets were used in this project; jazz midi data set, classical music midi data set classified by composers, and Arabic music midi data set. All the data sets are included in the following \hyperlink{https://github.com/pratiknag/MusicGeneration_with_CAN}{github} repository. Both the jazz data set and the classical music data set were retrieved from \hyperlink{https://www.kaggle.com/}{kaggle}. The Arabic data set was collected manually and cleaned to cancel out some of the repetitive notes such as drums.

\subsection{Data Preprocessing}
\begin{figure}[h]
    \centering
    \includegraphics{ 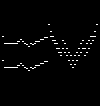}
    \caption{Sample of musical piece by the composer Albeniz in image format}
    \label{fig5}
\end{figure}

The data used in this project contain midi files. In order to pass them to the GAN or the CAN the midi files were converted to images by converting each note to a pixel location. The resulting image file can be converted back to midi file without losing any of the information (see Figure \ref{fig5}). 

\subsection{Music GAN Architecture}
The architecture used for the music GAN is presented here. Two networks were included in the architecture: a generator network and a discriminator network. A random vector $z$ was used as input by the generator to produce an image representing a MIDI file. The generated image was then passed to the discriminator, which attempted to distinguish between the real image dataset and the generated (fake) images. The GAN was trained on the three datasets separately, resulting in three distinct generators: one capable of producing jazz musical pieces, another capable of generating classical music pieces, and the third trained on manually collected and cleaned data to generate Arabic musical pieces.

\begin{figure}[h]
    \centering
    \includegraphics[width=\textwidth]{ 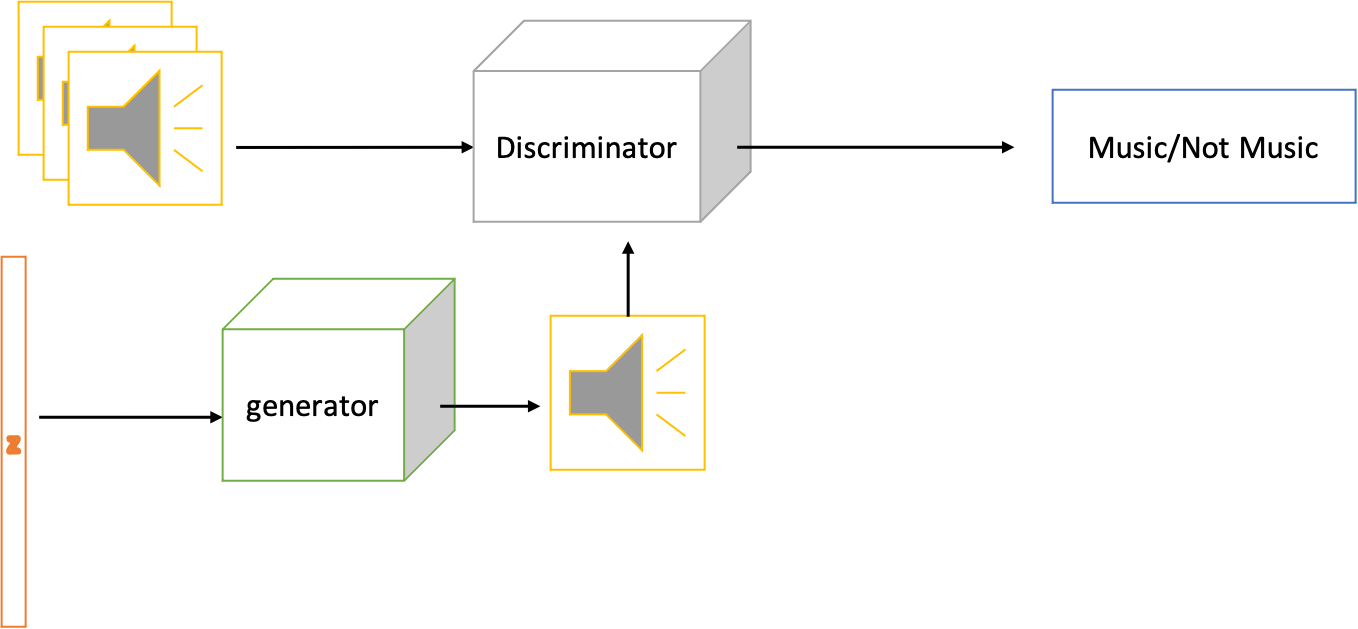}
    \caption{GAN architecture}
    \label{fig6}
\end{figure}

\begin{figure}[h]
    \centering
    \includegraphics[width=\textwidth]{ 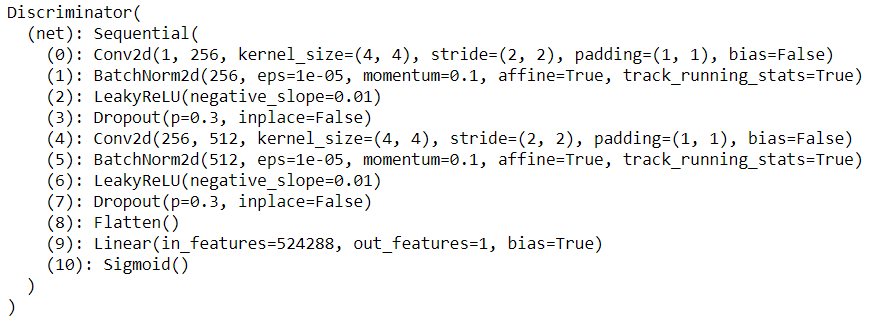}
    \caption{GAN Discriminator block}
    \label{fig7}
\end{figure}

\begin{figure}[h]
    \centering
    \includegraphics[width=\textwidth]{ 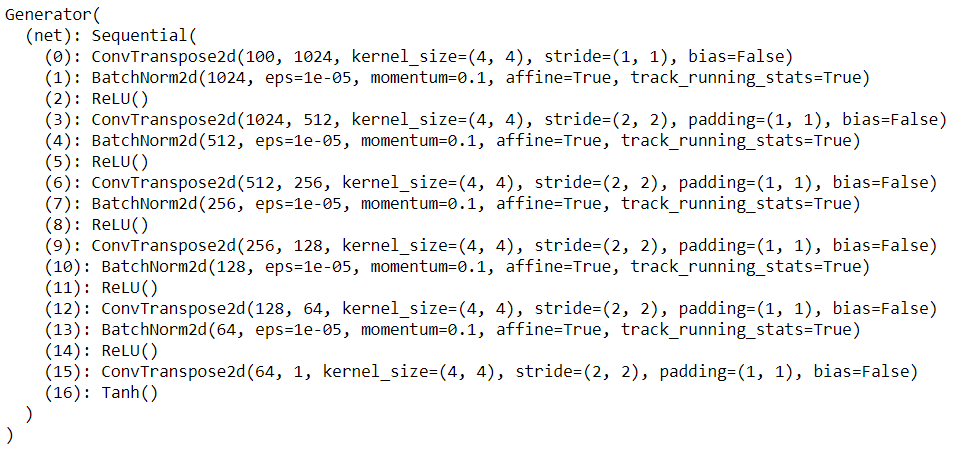}
    \caption{GAN Generator block}
    \label{fig8}
\end{figure}

A detailed architecture of Discriminator and Generator. \ref{fig7},\ref{fig8}

\subsection{Music CAN and Unrolled CAN Architecture}
In the music CAN the architecture \cite{sbai2018design,DBLP} is still made up of two neural networks that compete against each other. However, the loss here stresses on the generator motivating it to produce a musical piece that cannot be classified as belonging to one class. In other words, the discriminator should be capable of classifying the original real data to their composers or class label.  The discriminator should not be able to find a suitable class for the generated piece of music. Thus the input data must be labeled by the class it belongs to in order to reduce the loss on real data classification. The generator architecture is similar to that of the GAN where it takes a random z vector as an input and generates an image representing the midi file.
\begin{figure}[!h]
    \centering
    \includegraphics[width=\textwidth]{ 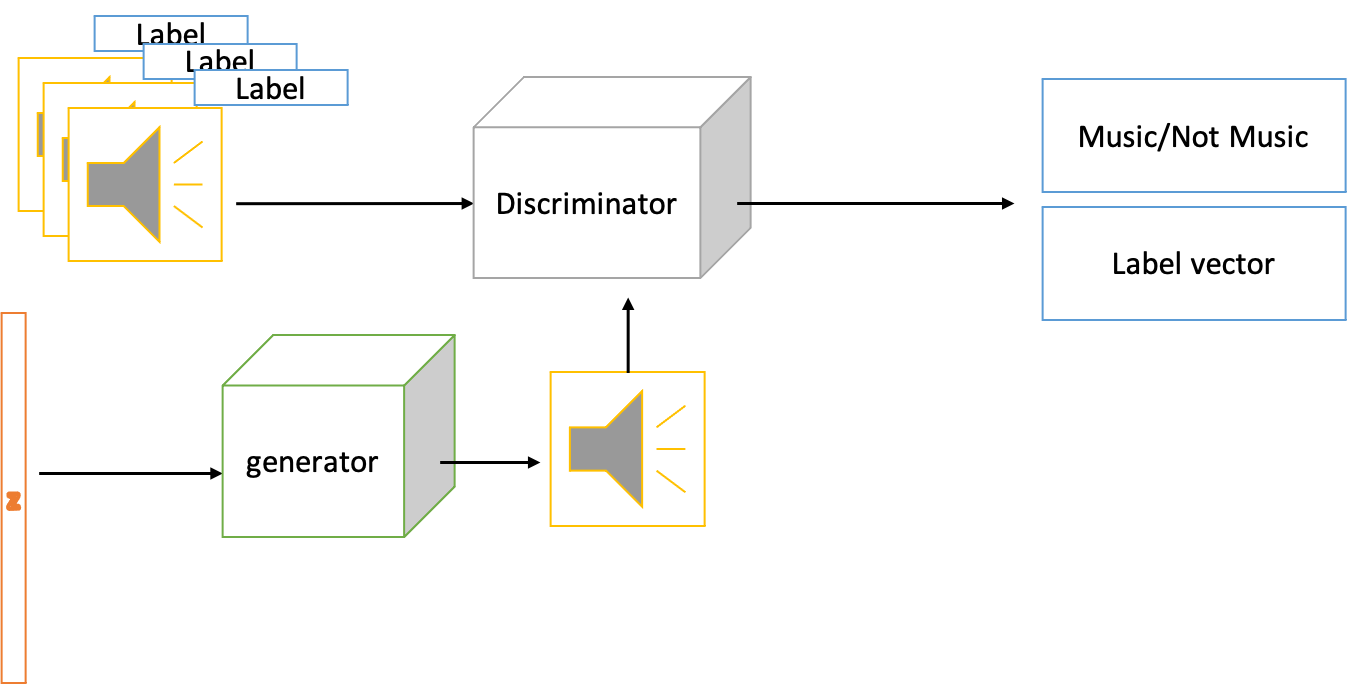}
    \caption{CAN architecture}
    \label{fig9}
\end{figure}
A detailed architecture of Discriminator and Generator is shown in Figure \ref{fig10} and \ref{fig11}.
\begin{figure}[h]
    \centering
    \includegraphics[width=\textwidth]{ 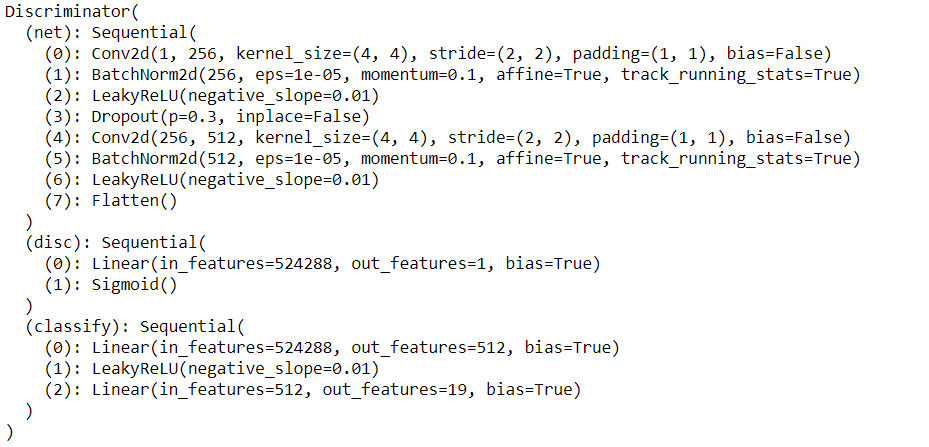}
    \caption{CAN Discriminator block}
    \label{fig10}
\end{figure}
\begin{figure}[h]
    \centering
    \includegraphics[width=\textwidth]{ 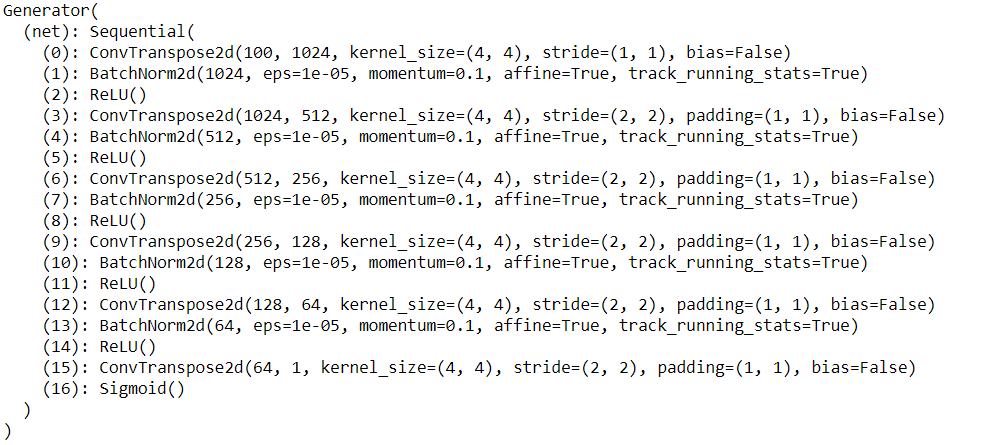}
    \caption{CAN Generator block}
    \label{fig11}
\end{figure}

For the CAN generator architecture, a size 100 random \(z\) vector is converted into a \(512 \times 4 \times 4\) tensor through the use of convolution and batch normalization layers. Convolution and batch normalization are then applied sequentially to produce a \(256 \times 8 \times 8\) tensor, followed by a \(128 \times 32 \times 32\) tensor, and subsequently a \(64 \times 64 \times 64\) tensor. Finally, convolution and batch normalization are applied once more to generate a \(1 \times 128 \times 128\) tensor, which represents the graphical form of the MIDI file.

For the discriminator, two convolution layers are employed with batch normalization applied to both, followed by a leaky ReLU activation and a fully connected layer.

\section{Results}

This section will discuss and compare the results of the three architectures; GAN, CAN, and unrolled CAN. The results are presented after post-processing the images and converting them to musical notes extracted from their corresponding midi files. 

\subsection{Scoring Novelty Generated Musical Pieces}
\begin{figure}[!h]
    \centering
    \includegraphics[width=\textwidth]{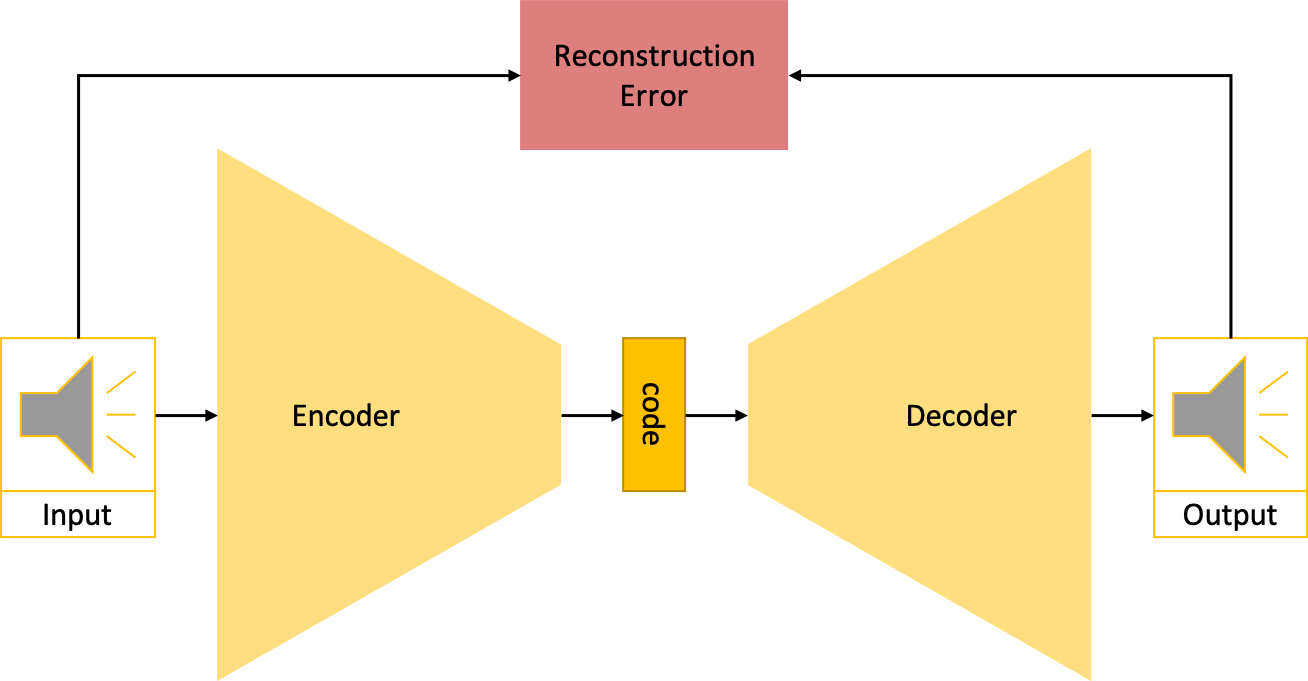}
    \caption{Auto-encoder network architecture}
    \label{fig14}
\end{figure}

In this paper, the expert gate technique \cite{aljundi2017expert} was followed, where an auto-encoder was trained on the reconstruction of the real dataset. The expert gate was trained to become proficient in reconstructing the real dataset, which was then used to evaluate the novelty of the generated images. The Mean Squared Error (MSE) was used as the scoring metric. Higher scores indicated greater creativity and deviation from the training set. After the training process, the reconstruction error of the real dataset was observed to have an average of 0.01 MSE. The network architecture is shown in Figure~\ref{fig14}.

\subsection{Music GAN Results}
The music GAN was capable of generating musical pieces that sounded and scored closely to the dataset. The output shown below in Figure~\ref{fig15} represents the result of training on the Jazz dataset. The generated data files exhibited very low reconstruction error, with an average of 0.02 MSE. However, the generated music produced by the GAN architecture was only able to capture the treble clef and was incapable of producing the bass clef.

\begin{figure}[!h]
    \centering
    \includegraphics[width=\textwidth]{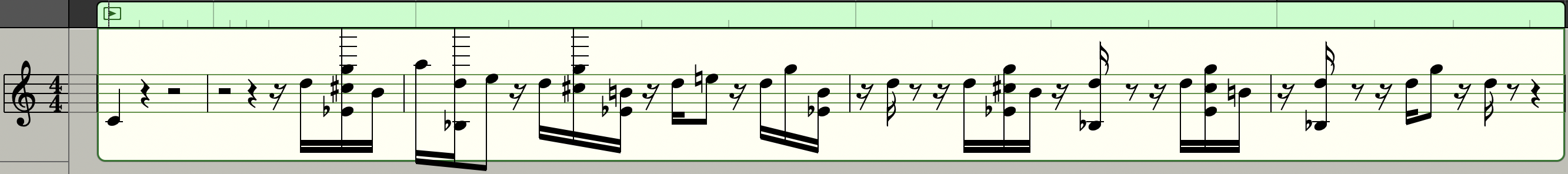}
    \caption{Output musical note from classical GAN architecture trained on the Jazz dataset}
    \label{fig15}
\end{figure}

\subsection{Music CAN Results}
\begin{figure}[!h]
    \centering
    \includegraphics[width=\textwidth]{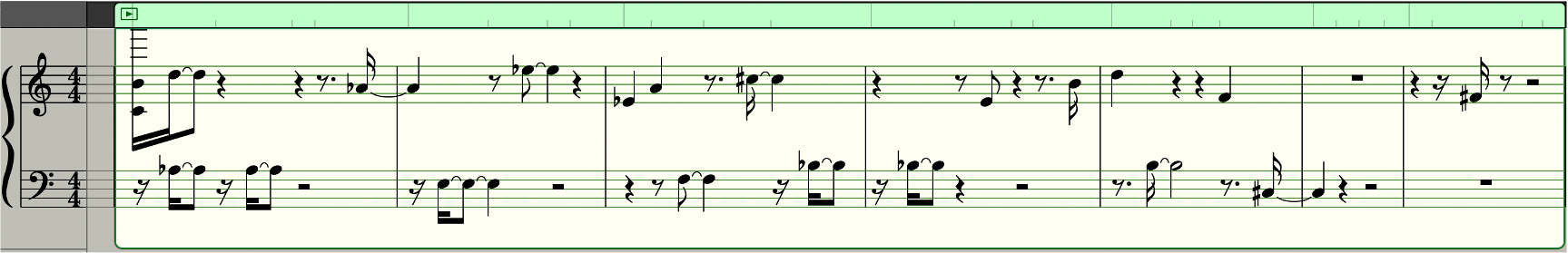}
    \caption{Output musical note from CAN architecture trained on the classical music dataset}
    \label{fig16}
\end{figure}

The CAN network was trained on the classical music dataset, where each composer was assigned as a class label. The goal was to generate new classical music that deviates from the styles of all the classical music composers. From the results shown in Figure~\ref{fig16}, an improvement over the GAN network can be observed, as the bass clef was captured with greater coherence between the notes. The average score of the generated music set was found to be 0.013 MSE. This indicates that, although the generated music is novel and exhibits high entropy between composers' styles, it remains relatively close to the original dataset.

\subsection{Unrolled Music CAN Results}
\begin{figure}[!h]
    \centering
    \includegraphics[width=\textwidth]{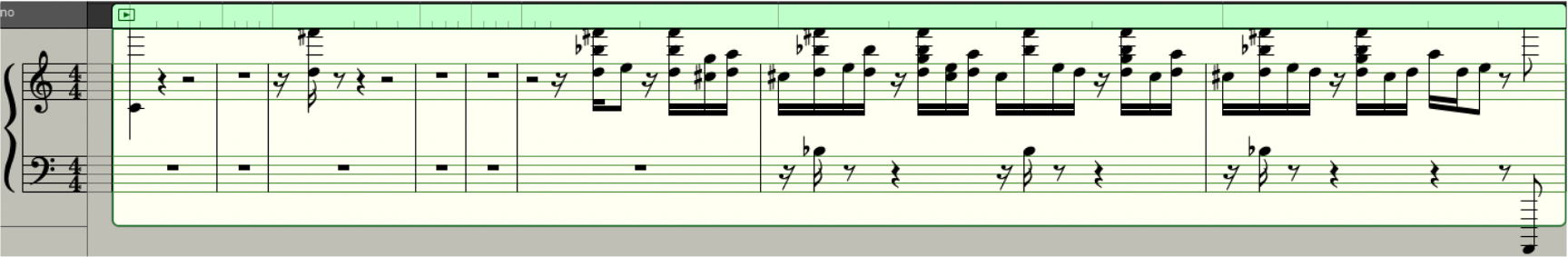}
    \caption{Output musical note from Unrolled CAN architecture trained on the classical music dataset}
    \label{fig17}
\end{figure}

The unrolled CAN architecture combined the strengths of both GAN and CAN, resulting in higher scores while maintaining high entropy between composers' styles. The average score of the music generated by the unrolled CAN was observed to be 0.043 MSE. This improvement is highlighted in Figure~\ref{fig17}.

\subsection{Discussion}

It is observed that unrolled CAN is capable of deviating from learned styles and also deviating from the data set while still being able to trick the discriminator network. 

\begin{table}[h]
\begin{center}
\begin{tabular}{|l|c|}
\hline
Network & Max Novelty Score \\
\hline\hline
GAN &  0.029\\
CAN & 0.013 \\
Unrolled CAN & \textbf{0.043}\\
\hline
\end{tabular}
\end{center}
\caption{Results for different comparing models.}
\end{table}
None the less the CAN architecture was very good at generating music that deviates from the composers' styles but converging to the generation of similar results leading to mode collapse which could be the reason why the average reconstruction error was very low.

\subsection{Reproducing Results}
All the codes related to this work have been uploaded to the following GitHub repository: \href{https://github.com/pratiknag/MusicGeneration_with_CAN}{GITHUB REPO}. \par
To train the Unrolled CAN model on the dataset, the file \texttt{GAN\_reconstructed.- ipynb} located in the folder \texttt{CAN\_models} should be executed. The files \texttt{training.py} and \texttt{model.py} contain all the model-related configurations. MIDI datasets can be preprocessed into images using the \texttt{data\_pre\_processing.ipynb} file from the home directory. \par

To train the second-best model, the DCGAN, the folder \texttt{GAN\_models} should be accessed, and the file \texttt{GAN-Copy2.ipynb} must be run. Before execution, the proper folder path containing the image representation of the MIDI datasets should be specified in the \texttt{GAN-Copy2.ipynb} file. Some generated results for both models have been uploaded to their respective folders. \par

The \texttt{novelty\_score.ipynb} file, placed in the home directory, has been utilized to calculate the novelty score, as discussed in \cite{aljundi2017expert}. This file is located in the home directory. \par

The datasets used in this work are publicly available:  
\href{https://www.kaggle.com/saikayala/jazz-ml-ready-midi}{Jazz MIDI Dataset} and \href{https://www.kaggle.com/soumikrakshit/classical-music-midi}{Classical MIDI Dataset}.

\section{Conclusion}
This marks a new artistic era where machines are introduced into the creative domain. Machines are now capable of contributing to the artistic and imaginative sectors, areas where their involvement was previously limited. It is encouraging to witness machines evolving in artificial intelligence and generating ideas that humans might find challenging to perceive. While humans require years to master music theory and play fluently, the algorithm presented in this paper demonstrates the ability to learn musical instruments fluently within minutes and produce music in seconds. \par

This paper explored different ways musical styles are classified and reviewed multiple architectures. It was demonstrated that the GAN structure \cite{goodfellow2014generative} and CAN \cite{DBLP} are fully capable of generating music. Additionally, a novel modification to the CAN was introduced by unrolling it, as detailed in \cite{metz2017unrolled}. This project combined art and science, leveraging machine learning to create novel and innovative music. Such advancements could inspire new playlists for hotels, aid singers and music composers in seeking new inspirations, and reaffirm the purpose of machines as tools to benefit humanity. \par

Future improvements could include expanding the dataset to incorporate a wider variety of musical styles and introducing multiple instruments within a single piece. This would enable the development of an unrolled concert CAN capable of orchestrating complex compositions using multiple instruments.

\bibliographystyle{splncs04}
\bibliography{egbib}
%





\end{document}